\documentclass[aps,prd,preprint,showpacs,nofootinbib,superscriptaddress]{revtex4}
\usepackage{verbatim}
\usepackage{slashed}
\usepackage{epsfig}
\usepackage{feynmf}
\usepackage{graphicx}
\usepackage{amsmath, amssymb, amsfonts}
\usepackage{mathrsfs}
\usepackage{bm}
\usepackage{relsize}
\def\babar{\mbox{\slshape B\kern-0.1em{\smaller A}\kern-0.1em
B\kern-0.1em{\smaller A\kern-0.2em R}}}

\begin{document}

\title{\bf B-meson Semi-inclusive Decay to $2^{-+}$ Charmonium
in NRQCD and X(3872)}


\author{Ying Fan$^1$\footnote{Current address:
Department of Physics, Korea University, Seoul 136-701, Korea.},
~Jin-Zhao Li$^{1}$,
~Ce Meng$^{1}$,
~and Kuang-Ta Chao$^{1,2}$
\\[2mm]
{\it\small$\rm ^1$Department of Physics and State Key Laboratory of Nuclear Physics and Technology,}\\
{\it\small Peking University, Beijing 100871, China}\\
{\it\small$\rm ^2$Center for High Energy Physics, Peking University,
Beijing 100871, China} }
\noaffiliation

\begin{abstract}
The semi-inclusive B-meson decay into spin-singlet D-wave $2^{-+}$
charmonium, $B\to\eta_{c2}+X$, is studied in non-relativistic QCD
(NRQCD). Both color-singlet and color-octet contributions are
calculated at next-to-leading order (NLO) in the strong coupling
constant $\alpha_s$. The non-perturbative long-distance matrix
elements are evaluated using operator evolution equations. It is
found that the color-singlet $^1D_2$ contribution is tiny, while the
color-octet channels make dominant contributions. The estimated
branching ratio $B(B\to\eta_{c2}+X)$ is about $0.41\,\times10^{-4}$
in the Naive Dimensional Regularization (NDR) scheme and
$1.24\,\times10^{-4}$ in the t'Hooft-Veltman (HV) scheme, with
renormalization scale $\mu=m_b=4.8$\,GeV. The scheme-sensitivity of
these numerical results is due to cancelation between
${}^1S_0^{[8]}$ and ${}^1P_1^{[8]}$ contributions. The
$\mu$-dependence curves of NLO branching ratios in both schemes are
also shown, with $\mu$ varying from $\frac{m_b}{2}$ to $2m_b$ and
the NRQCD factorization or renormalization scale $\mu_{\Lambda}$
taken to be $2m_c$. Comparison of the estimated branching ratio of
$B\to\eta_{c2}+X$ with the observed branching ratio of $B \to
X(3872)+K$ may lead to the conclusion that $X(3872)$ is unlikely to
be the $2^{-+}$ charmonium state $\eta_{c2}$.
\end{abstract}

\maketitle

\section{introduction}
One of the missing states in the charmonium family, the
$\eta_{c2}({}^1D_2)$, is the only missing spin-singlet low-lying
D-wave charmonium state. Its mass is predicted to be within 3.80 to
3.84 GeV \cite{Eichten:2004uh,Barnes:2005pb,Li:2009zu}, which lies
between the $D\bar{D}$ and the $D^*\bar{D}$ thresholds. The $J^{PC}$
quantum number of $\eta_{c2}$ is $2^{-+}$, thus its decay to
$D\bar{D}$ is forbidden. Therefore, this is a narrow resonance
state, and its main decay modes are the electromagnetic and hadronic
transitions to lower-lying S-, P-wave charmonium states and the
annihilation decays to light hadrons. Previously, we calculated the
inclusive light hadronic decay width of the ${}^1D_2$ state at
next-to-leading order (NLO) in $\alpha_s$ \cite{Fan:2009cj} in the
framework of non-relativistic QCD (NRQCD). The results show that
with the total width of $\eta_{c2}$ estimated to be about 660-810
keV, the branching ratio of the electric dipole transition
$\eta_{c2}\rightarrow\gamma h_c$ is about $(44-54)\%$, which will be
useful in searching for this missing charmonium state through
$\eta_{c2}\rightarrow\gamma h_c$ followed by
$h_c\rightarrow\gamma\eta_c$ and other processes.

The NRQCD factorization method\cite{Bodwin:1994jh} was adopted in
our calculation of $\eta_{c2}$ light hadronic decay. Within this
framework, the inclusive decay and production of heavy quarkonium
can be factorized into two parts, the short-distance coefficients
and the long-distance matrix elements. A color-octet heavy quark and
anti-quark pair annihilated or produced at short distances can
evolve into a color-singlet heavy quarkonium at long distances via
electric or magnetic transitions by emitting soft gluons,
This color-octet mechanism has been used to remove the infrared (IR)
divergences in
P-wave
\cite{Bodwin:1994jh,Huang:1996fa,Huang:1996sw,Huang:1996cs,Petrelli:1997ge,Maltoni:1999phd}
and D-wave \cite{He:2008xb,Fan:2009cj,He:2009bf} charmonium decays.

Now, we turn to the B-meson non-leptonic decays, which have played
an important role in discovering new resonances, especially new
charmonium and charmonium-like states in recent years.  The
branching fractions of B-meson inclusive decays into S-wave and
P-wave charmonia, of $\mathcal{O}(10^{-3})$ to
$\mathcal{O}(10^{-2})$\cite{Nakamura:2010zzi}, are relatively large.
Therefore, we may also expect to search for D-wave charmonia in
B-meson decays, and in particular to search for the spin-singlet
D-wave charmonium $\eta_{c2}$ in $B\rightarrow\eta_{c2}+X$. Like the
charmonium light hadronic decay, charmonium production in B-meson
semi-inclusive decay may also be factorized in NRQCD as
\begin{equation}\label{eq:factorization}
\Gamma(B\rightarrow H+X)=\sum_n C(b\rightarrow
c\bar{c}[n]+X)\langle\mathcal{O}^H [n]\rangle,
\end{equation}
where the sum runs over all contributing Fock states. The
short-distance coefficients $C(b\rightarrow c\bar{c}[n]+X)$ can be
perturbatively calculated up to any order in $\alpha_s$; while the
long-distance matrix elements $\langle\mathcal{O}^H [n]\rangle$
should be determined non-perturbatively.
One may refer to \cite{Beneke:1998ks,Maltoni:1999phd} for more
discussions on the feasibility of Eq.~(\ref{eq:factorization}).

S-wave and P-wave charmonium production in B-meson semi-inclusive
decays have already been studied by many authors in the literature
\cite{Bergstrom:1994vc,Ko:1995iv,Fleming:1996pt,Soares:1997ir,Beneke:1998ks,Maltoni:1999phd}.
In \cite{Beneke:1998ks,Maltoni:1999phd}, it was found  that the
experimentally observed branching fractions for $J/\psi$ and
$\psi^\prime$ could be accounted for by NLO calculations, while for
$\chi_{c1}$ and $\chi_{c2}$ the branching ratios were still
difficult to explain.   In \cite{Yuan:1997we}, the branching
fractions for D-wave charmonium production in B-meson semi-inclusive
decays were calculated to be of $\mathcal{O}(10^{-3})$ in NRQCD at
leading order (LO), where the NRQCD velocity scaling rules were used
to estimate the long-distance matrix elements. Similar results but
somewhat larger branching fractions were also obtained in \cite{Ko}.
However, the NLO QCD corrections are found to be very important in
many heavy quarkonium production processes, e.g. in $e^+e^-$
annihilation\cite{NLOee}, hadroproduction\cite{NLOhad1,NLOhad8}, and
photoproduction\cite{NLOpho}. Moreover, the velocity scaling rules
are too rough to give a quantitative estimate for the long-distance
matrix elements. Therefore, for D-wave charmonium production in
B-meson semi-inclusive decays, aside from \cite{Yuan:1997we,Ko}, a
NLO calculation and a better estimate for the matrix elements are
necessary.

Another important motivation for carrying out this study concerns
the long-standing puzzle of the nature of $X(3872)$. Previous
studies assumed that the quantum numbers of the $X(3872)$ were
$J^{PC}=1^{++}$, and this was supported by a number of measurements.
However, the new \babar~measurement of $X(3872)\to
J/\psi\pi^+\pi^-\pi^0$~\cite{delAmoSanchez:2010jr} favors the
negative-parity assignment $2^{-+}$. Nevertheless, people still
argue that the observed properties of $X(3872)$ strongly disfavor
the $2^{-+}$
assignment\cite{Jia:2010jn,Burns:2010qq,Kalashnikova:2010hv,Ke:2011}.
Recently, \cite{Mehen:2011ds} proposed that the angular
distributions of decay products could be used to distinguish between
the $1^{++}$ and $2^{-+}$ assignments of $X(3872)$. In this paper,
we will further clarify this problem by calculating the ${}^1D_2$
charmonium production rate in B-meson semi-inclusive decay. We will
compare the calculated branching ratio $B\rightarrow \eta_{c2}+X$,
with the experimental measurement of $Br(B\rightarrow X(3872) K)$,
and then discuss if $X(3872)$ can be the $2^{-+}$ charmonium
$\eta_{c2}$.

The paper is organized as follows. In Sec.\,\ref{sec:LO} and
\ref{sec:NLO}, decay widths of four contributing Fock states at tree
and one-loop levels are calculated both in QCD and NRQCD, and finite
short-distance coefficients $C(b\rightarrow c\bar{c}[n]+X)$ for
different components $c\bar{c}[n]$ are obtained respectively after
matching between QCD and NRQCD. Computation methods adopted in real
and virtual corrections are discussed too. The long-distance matrix
elements are estimated using operator evolution equations. In
Sec.\,\ref{sec:results}, numerical results are given and analyzed.
And finally the possibility of assigning the $\eta_{c2}$ as X(3872)
is discussed.

\section{Leading-order (LO) contribution}
\label{sec:LO}

We use the same description as in
\cite{Maltoni:1999phd,Beneke:1998ks}. The weak decay $b\rightarrow
c\bar{c}+ s/d$ occurs at energy scales much lower than the W boson
mass $m_W$. Integrating out the hard scale and making Fierz
transformation, we finally arrive at the effective Hamiltonian
\begin{equation}\label{eq:Heff}
H_{eff}=\frac{G_F}{\sqrt{2}}\sum_{q=s,d}\left\{V_{cb}^*V_{cq} \left[\frac{1}{3}C_{[1]}
(\mu)\mathcal{O}_1(\mu)+C_{[8]}(\mu)\mathcal{O}_8(\mu) \right]-V_{tb}^*V_{tq}\sum_{i=3}^6
C_i(\mu)\mathcal{O}_i(\mu)\right\},
\end{equation}
where the $c\bar{c}$ pair is either in a color singlet or a color
octet configuration, denoted by $\mathcal{O}_1$ and $\mathcal{O}_8$
respectively,
\begin{eqnarray}
\mathcal{O}_1&=&[\bar{c}\gamma_\mu(1-\gamma_5)c][\bar{b}\gamma^\mu(1-\gamma_5)q],\nonumber\\
\mathcal{O}_8&=&[\bar{c}T^a\gamma_\mu(1-\gamma_5)c][\bar{b}T^a\gamma^\mu(1-\gamma_5)q].
\end{eqnarray}
$\mathcal{O}_{3-6}$ are the QCD penguin operators
\cite{Buchalla:1995vs}. $C_{[1]}(\mu)$ and $C_{[8]}(\mu)$ are the
Wilson coefficients of $\mathcal{O}_1$ and $\mathcal{O}_8$, and
related to another group of coefficients $C_{+}(\mu)$ and
$C_{-}(\mu)$ through
\begin{eqnarray}
C_{[1]}(\mu) &=& 2 C_+(\mu) - C_-(\mu),
\nonumber\\
C_{[8]}(\mu) &=&   C_+(\mu) + C_-(\mu).
\end{eqnarray}
At LO, expressions for $C_\pm(\mu)$ are
\begin{equation}
C^{LO}_\pm(\mu) = \left[ \frac{\alpha^{LO}_s(m_W)}{\alpha^{LO}_s(\mu)} \right]
^{\gamma^{(0)}_\pm/(2 \beta_0)},
\end{equation}
with the one-loop anomalous dimension
\begin{equation}
\gamma^{(0)}_\pm = \pm \,2 \,(3 \mp 1),
\end{equation}
and $\alpha_s$
\begin{equation}
\alpha^{LO}_s(\mu) = \frac{4 \pi}{\beta_0 \ln [\mu^2/(\Lambda^{LO}_{QCD})^2]},
\end{equation}
where $\beta_0 = 11 - \frac{2}{3} N_f $. We choose
$m_W=80.399$\,GeV\cite{Nakamura:2010zzi}, $m_Z=91.1876$\,GeV,
$m_b=4.8$ GeV, $N_f=4$, and $\Lambda^{LO}_{QCD}=128$ MeV for four
flavors to adjust $\alpha_s(m_Z)$ to be 0.119 for five flavors.

Only four configurations contribute to $\eta_{c2}$ production at LO
in $v$, the velocity of heavy quark or anti-quark in charmonium rest
frame:
\begin{equation}\label{expansion:Fock}
|\eta_{c2}\rangle=\mathcal{O}(1)|{}^1D_2^{[1]}
\rangle+\mathcal{O}(v)|{}^1P_1^{[8]}g\rangle+\mathcal{O}(v^2)|{}^1S_0^{[1,8]}gg\rangle+\cdots.
\end{equation}
With the Fock state expansion Eq.~(\ref{expansion:Fock}),
we have
\begin{eqnarray}
\Gamma(b\rightarrow \eta_{c2}X)&=&\Gamma(b\rightarrow {}^1S_0^{[1]} X)+
\Gamma(b\rightarrow {}^1S_0^{[8]} X)+\Gamma(b\rightarrow {}^1P_1^{[8]} X)+
\Gamma(b\rightarrow {}^1D_2^{[1]} X)\nonumber\\
&=&C({}^1S_0^{[1]})\langle\mathcal{O}_1({}^1S_0)\rangle+ C({}^1S_0^{[8]})
\langle\mathcal{O}_8({}^1S_0)\rangle+ C({}^1P_1^{[8]})
\frac{\langle\mathcal{O}_8({}^1P_1)\rangle}{m_c^2}+ C({}^1D_2^{[1]})
\frac{\langle\mathcal{O}_1({}^1D_2)\rangle}{m_c^4}.\nonumber\\
\end{eqnarray}
$\langle\mathcal{O}_1({}^1S_0)\rangle$,
$\langle\mathcal{O}_8({}^1S_0)\rangle$,
$\langle\mathcal{O}_{8}(^1P_{1})\rangle$ and
$\langle\mathcal{O}_{1}(^{1}D_{2})\rangle$ are the production matrix
elements of four-fermion operators defined in
\cite{Bodwin:1994jh,Braaten:2002fi}:
\begin{eqnarray}
\mathcal{O}_{1}(^1S_{0})&=&\chi^{\dagger}\psi\left(a_H^\dagger a_H\right)\psi^{\dagger}\chi,\nonumber\\
\mathcal{O}_{8}(^1S_{0})&=&\chi^{\dagger}T^{a}\psi\left(a_H^\dagger a_H\right)\psi^{\dagger}T^{a}\chi,
\nonumber\\
\mathcal{O}_{8}(^1P_{1})&=&\chi^{\dagger}(-\frac{i}{2}
\overleftrightarrow{\boldsymbol{D}})T^{a}\psi\left(a_H^\dagger a_H\right)\cdot
\psi^{\dagger}(-\frac{i}{2}\overleftrightarrow{\boldsymbol{D}})T^{a}\chi,\nonumber\\
\mathcal{O}_{1}(^{1}D_{2})&=&\chi^{\dagger}S^{ij}\psi\left(a_H^\dagger a_H\right)
\psi^{\dagger}S^{ij}\chi, \end{eqnarray} where
$\overleftrightarrow{\boldsymbol{D}}=\overrightarrow{\boldsymbol{D}}-\overleftarrow{\boldsymbol{D}}$
and $S^{ij}=(-\frac{i}{2})^{2}
(\overleftrightarrow{D}^{i}\overleftrightarrow{D}^{j}-\frac{1}{3}\overleftrightarrow{\boldsymbol
D}^{2}\delta^{ij})$.

\begin{figure}[htbp]
\begin{center}
\includegraphics[scale=0.8]{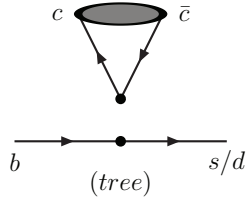}
\caption{LO Feynman diagram of $b\rightarrow c\bar{c}+X$.}
\label{figure:LO}
\end{center}
\end{figure}

\begin{figure}[htbp]
\begin{center}
\includegraphics[scale=0.55]{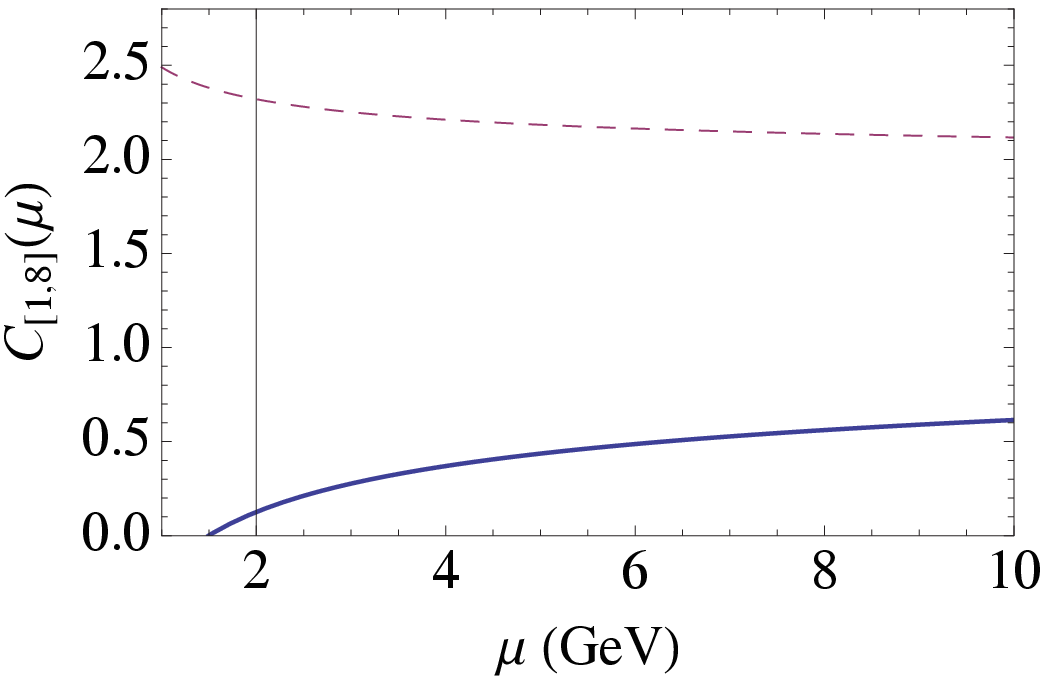}
%
\hspace{1cm}
\includegraphics[scale= 0.55]{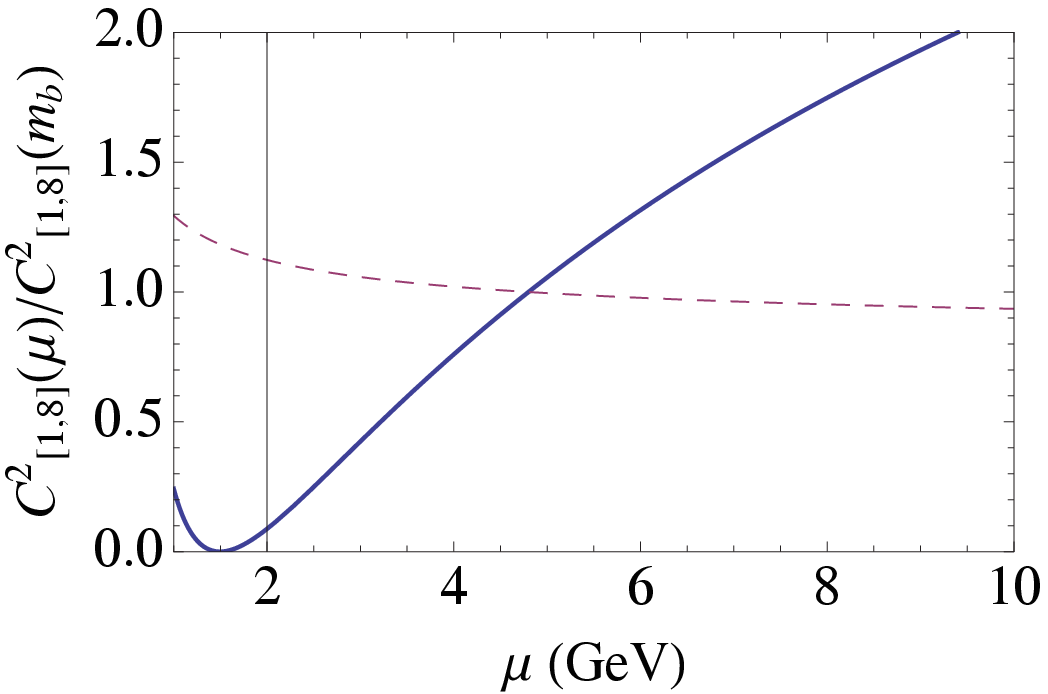}
\caption{LO $\mu$-dependence curves of $C_{[1,8]}(\mu)$. The solid
line denotes $C_{[1]}(\mu)$, and the dashed line $C_{[8]}(\mu)$.
Ratios of $C_{[1,8]}^2(\mu)$/$C_{[1,8]}^2(m_b)$ as functions of
$\mu$ are also shown.} \label{figure:LOC18&LOC18square}
\end{center}
\end{figure}

We use Wolfram Mathematica 7.0.1.0, feynarts-3.4, and FeynCalc 6.0.
At tree-level, the coupling vertex structure
$\bar{c}\gamma_\mu(1-\gamma_5)c$ restricts possible $J^{PC}$ numbers
of charmonium states. Matching amplitudes in both QCD and NRQCD at
LO leads to finite short-distance coefficients
\begin{eqnarray}\label{width:LO}
C( {}^1S_0^{[1]} )&=& \Gamma_0C_{[1]}^2 3(1-\eta)^2,\nonumber\\
C( {}^1S_0^{[8]} )&=& \Gamma_0C_{[8]}^2 \frac{9}{2}(1-\eta)^2,\nonumber\\
C( {}^1P_1^{[8]} )&=& 0,\nonumber\\
C( {}^1D_2^{[1]} )&=& 0,
\end{eqnarray}
where
\begin{equation}
\Gamma_0=\frac{G_F^2|V_{bc}|^2m_b^3}{216\pi(2m_c)},\hspace{1cm} \eta=\frac{4m_c^2}{m_b^2},
\end{equation}
and $|V_{cs}|^2+|V_{cd}|^2\approx 1$ has been used. For the LO
Feynman diagram, see Fig.~[\ref{figure:LO}].
The strong dependence on renormalization scale $\mu$ of
$C^2_{[1,8]}(\mu)$ at LO causes the results in Eq.~(\ref{width:LO})
unreliable (see Fig.~[\ref{figure:LOC18&LOC18square}]) and calls for
higher order corrections.
The QCD Penguin operators in Eq.~(\ref{eq:Heff}) also contribute to
non-zero tree-level decay width, although their contribution is tiny
due to the smallness of $C_{3-6}(\mu)$. We will neglect their
$\mu$-dependence and adopt those values given in
\cite{Beneke:1998ks,Maltoni:1999phd}, for they chose the same values
for $m_b$, $\Lambda_{QCD}^{LO}$ as ours. $C_3(m_b)=0.010$,
$C_4(m_b)=-0.024$,  $C_5(m_b)=0.007$ and $C_6(m_b)=-0.028$. Together
with $C^{LO}_{[1]}(m_b)=0.42$ and $C^{LO}_{[8]}(m_b)=2.19$, the
Penguin contribution is
\begin{eqnarray}
\delta_P[{}^1S_0^{[1]}] &=& 2\frac{3 (C_3-C_5)+C_4-C_6}{C^{LO}_{[1]}}\nonumber\\
&=&0.06,\nonumber\\
\delta_P[{}^1S_0^{[8]}] &=& 4\frac{C_4-C_6}{C^{LO}_{[8]}}\nonumber\\
&=&0.007,
\end{eqnarray}
which add corrections to tree-level short-distance coefficients in Eq.~(\ref{width:LO})
\begin{eqnarray}\label{width:LOwithPenguin}
C( {}^1S_0^{[1]} )&=& \Gamma_0C_{[1]}^2 3(1-\eta)^2(1+\delta_P[{}^1S_0^{[1]}]),\nonumber\\
C( {}^1S_0^{[8]} )&=& \Gamma_0C_{[8]}^2 \frac{9}{2}(1-\eta)^2(1+\delta_P[{}^1S_0^{[8]}]),\nonumber\\
C( {}^1P_1^{[8]} )&=& 0,\nonumber\\
C( {}^1D_2^{[1]} )&=& 0.
\end{eqnarray}

\section{NLO calculation and divergence cancellation}
\label{sec:NLO}

\subsection{Real Corrections}

\begin{figure}[htbp]
\begin{center}
\includegraphics[scale=0.8]{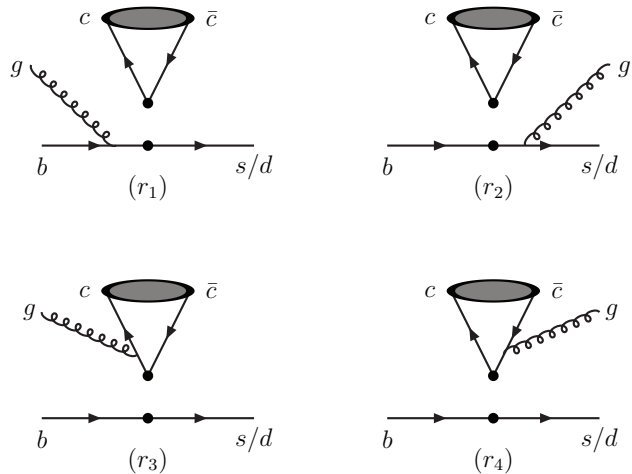}
\caption{Real correction Feynman diagrams of $b\rightarrow
c\bar{c}+X$.} \label{figure:real}
\end{center}
\end{figure}

Gluon mass regularization is adopted in our calculation, therefore
$\gamma_5$ matrix can be treated in 4-dimension. Real correction
figures are in Fig.~[\ref{figure:real}].
Divergences are separated from the finite parts in the amplitude
squared. Two kinds of divergences appear: the soft and the
collinear. Three divergent regions exist: soft, soft-collinear and
hard-collinear. Take ${}^1S_0^{[1]}$ for example. In the soft
region, the  gluon connected to the incoming bottom quark turns
soft, {\it i.e.}, its momentum goes to zero (($r_1$) of
Fig.~[\ref{figure:real}]); in the soft-collinear region, b-quark
gluon turns soft and at the same time s/d-quark gluon is collinear
with the outgoing s/d quark, or their momenta are parallel to each
other (($r_1$) and ($r_2$) of Fig.~[\ref{figure:real}]); and in the
hard-collinear region, s/d-quark gluon runs parallel to the
s/d-quark (($r_2$) of Fig.~[\ref{figure:real}]). IR divergences in
($r_3$) and ($r_4$) of Fig.~[\ref{figure:real}] cancel each other.
We take the following parametrization
\begin{equation}
b(p_1)\rightarrow c(p_4) + \bar{c}(p_3) + s/d(p_5) + g(p_6),
\end{equation}
and the quark propagators in four quark lines have denominators
\begin{eqnarray}
N_1&\equiv&-2p_1\cdot p_6 + p_6^2\,,\nonumber\\
N_4&\equiv&2p_4\cdot p_6 + p_6^2\,,\nonumber\\
N_3&\equiv&2p_3\cdot p_6 + p_6^2\,,\nonumber\\
N_5&\equiv&2p_5\cdot p_6 + p_6^2\,,
\end{eqnarray}
respectively. For ${}^1S_0^{[1]}$, $p_3=p_4$ and $N_3=N_4$.
Divergent terms are extracted before doing phase space integration:
\begin{eqnarray}
\mbox{soft terms}&:&~~ \sim\frac{1}{N_1^2},\nonumber\\
\mbox{soft-collinear terms}&:&~~ \sim\frac{1}{N_1N_5},\nonumber\\
\mbox{hard-collinear terms}&:&~~ \sim\frac{1}{N_5},~\sim\frac{1}{N_5^2}.
\end{eqnarray}
Some of the hard-collinear terms are seemingly divergent but finally
contribute to the finite parts. The Mandelstam variables are
\begin{eqnarray}
s&=&(p_1-p_6)^2,\nonumber\\
t&=&(p_5+p_6)^2,\nonumber\\
u&=&(p_1-p_5)^2,
\end{eqnarray}
and
\begin{equation}
u=4m_c^2+m_b^2+\lambda^2-s-t,
\end{equation}
with $\lambda$ the non-zero gluon mass. Rescaling all the
dimensional variables with respect to $m_b$
\begin{eqnarray}
m_c&=&\frac{m_b}{2}\sqrt{\eta},\nonumber\\
\lambda&=&m_b\sqrt{\xi},\nonumber\\
\end{eqnarray}
and
\begin{eqnarray}\label{rescaling:st}
s&=&m_b^2(1-y+\xi),\nonumber\\
t&=&m_b^2(1-x+\eta),
\end{eqnarray}
we finally arrive at the amplitude squared expressed using
dimensionless variables $x$, $y$ instead of $s$ and $t$. Upper and
lower limits of $x$ and $y$ are derived from those of $s$ and $t$
via Eq.~(\ref{rescaling:st})
\begin{eqnarray}
y_{max}&=&1+\xi-\frac{1}{4(1+\eta-x)}(2\eta-x+\sqrt{x^2-4\eta})(-2+2\xi+x-\sqrt{x^2-4\eta})\,,\nonumber\\
y_{min}&=&1+\xi-\frac{1}{4(1+\eta-x)}(2\eta-x-\sqrt{x^2-4\eta})(-2+2\xi+x+\sqrt{x^2-4\eta})\,,\nonumber\\
x_{max}&=&1-\xi+\eta\,,\nonumber\\
x_{min}&=&2\sqrt{\eta}\,.
\end{eqnarray}
Phase space integration over $x$ is a little bit complicated,
and the Euler transformation is needed by introducing a new integration variable
\begin{equation}
\text{tt}\equiv\sqrt{\frac{x-2 \sqrt{\eta }}{x+2 \sqrt{\eta }}}
\end{equation}
to replace $x$ and its integration limits
\begin{eqnarray}
tt_{max}&=&\sqrt{\frac{\eta -2 \sqrt{\eta }-\xi +1}{\eta +2 \sqrt{\eta }-\xi +1}}\,,\nonumber\\
tt_{min}&=&0\,.
\end{eqnarray}
Divergences in ($r_3$) and ($r_4$) of Fig.~[\ref{figure:real}] can
not cancel each other for ${}^1S_0^{[8]}$, which makes divergent
terms more complicated. They also produce the only IR pole, the
residual divergence in ${}^1P_1^{[8]}$, which can be cancelled by
absorption into the redefinitions of non-perturbative matrix
elements of $^1S_0^{[1]}$ and $^1S_0^{[8]}$ states. Furthermore,
there is no divergence in real correction of ${}^1D_2^{[1]}$.

\subsection{Virtual Corrections}
\begin{figure}[htbp]
\begin{center}
\includegraphics[scale=0.8]{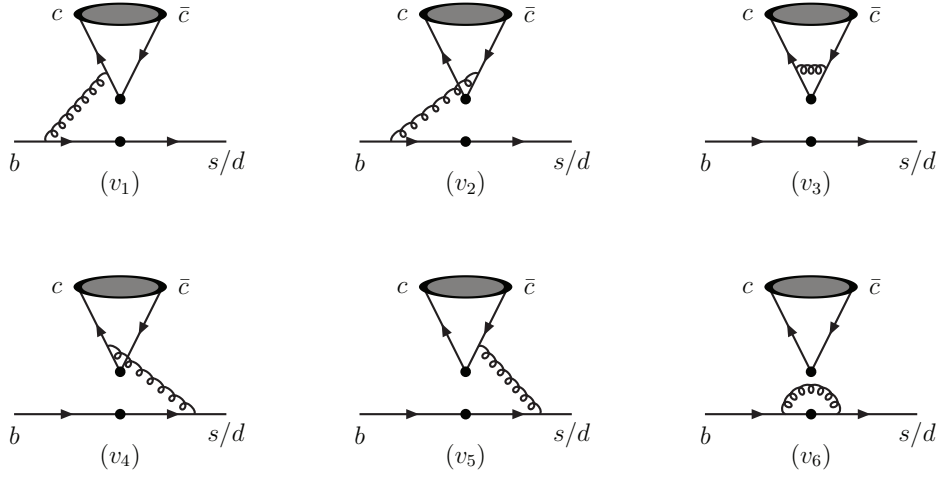}
\caption{Virtual correction Feynman diagrams of $b\rightarrow
c\bar{c}+X$.} \label{figure:virtual}
\end{center}
\end{figure}

In virtual corrections, IR divergences, soft and collinear, are
regulated with non-zero gluon mass like in real corrections.
Ultraviolet (UV) divergences are dimensionally regulated at the
amplitude level before projecting the free charm quark pair onto
certain charmonium bound state of particular angular momentum and
color. Virtual correction figures are in
Fig.~[\ref{figure:virtual}].
Each diagram in Fig.~[\ref{figure:virtual}] has an loop integration
over gluon momentum $q$. For example, in $(v_1)$ the UV divergent
loop integration has the form
\begin{equation}
\int\frac{d^Dq}{(2\pi)^D}\frac{q^\rho q^{\rho^\prime}}{(q^2-\lambda^2)((p_1-q)^2-m_b^2)((p_4-q)^2-m_c^2)},
\end{equation}
and the UV divergent term comes only from the region when $q\rightarrow \infty$
\begin{equation}
\int\frac{d^Dq}{(2\pi)^D}\frac{q^\rho q^{\rho^\prime}}{q^2\cdot q^2\cdot q^2},
\end{equation}
which is proportional to the D-dimensional metric tensor
$g^{\rho\rho^\prime}$. Thus corresponding fermion chain in $(v_1)$
reduces into
\begin{equation}
\Gamma_\mu\gamma_\rho\gamma_\alpha \otimes
\gamma^\alpha\gamma^\rho\Gamma^\mu.
\end{equation}
$\Gamma_\mu$ is the short form for electro-weak vertex $\gamma_\mu(1-\gamma_5)$.
UV divergent term extractions from structures like above are carried out
upon using the Fierz transformations
\begin{eqnarray}
\gamma_\rho\gamma_\alpha\Gamma_\mu \otimes
\gamma^\rho\gamma^\alpha\Gamma^\mu &=& (16+4X \epsilon_{UV}) \Gamma_\mu \otimes \Gamma^\mu+ E_X,
\nonumber\\
\Gamma_\mu\gamma_\rho\gamma_\alpha \otimes
\gamma^\alpha\gamma^\rho\Gamma^\mu &=& (4+4Y \epsilon_{UV}) \Gamma_\mu \otimes \Gamma^\mu+ E_Y,
\nonumber\\
\Gamma_\mu \otimes \gamma_\rho\gamma_\alpha\Gamma^\mu\gamma^\alpha\gamma^\rho
&=& (4+4Z \epsilon_{UV}) \Gamma_\mu \otimes \Gamma^\mu+ E_Z,
\end{eqnarray}
where the scheme dependence of $\gamma_5$ is fully extracted and
contained in scheme-dependent variables $X$, $Y$ and $Z$,
\begin{eqnarray}
\mbox{ NDR scheme} &:& ~~ X=-1,~~ Y=Z=-2\,;\nonumber\\
\mbox{ HV scheme} &:& ~~ X=-1,~~ Y=Z=0\,.
\end{eqnarray}
Hence, the $\gamma_5$ matrix in $\Gamma_\mu$ can still be kept in
4-dimension when evaluating the trace formalism. Evanescent
operators $E_X$, $E_Y$ and $E_Z$ exist only in $D\neq 4$ dimensions
but vanish in $D=4$ \cite{Buchalla:1995vs}. Therefore they make no
contribution to the decay widths, and can be discarded throughout
the calculations. Again for the ${}^1S_0^{[1]}$, self-energy
diagrams of $(v_3)$ and $(v_6)$ can only exist for color-singlet
electro-weak vertex, {\it i.e.}, only $C_{[1]}(\mu)$ appears. On the
contrary, the other four diagrams $(v_{1,2})$ and $(v_{4,5})$ can
only have $C_{[8]}(\mu)$ electro-weak vertex. Those six diagrams
only couple to the tree diagram with $C_{[1]}(\mu)$ vertex,
contributing to $C^2_{[1]}(\mu)$ and $C_{[1]}(\mu)C_{[8]}(\mu)$
terms, respectively. IR divergence of $(v_1)$ cancels that of
$(v_2)$, and $(v_4)$ cancels $(v_5)$.

\begin{figure}[htbp]
\begin{center}
\includegraphics[scale=0.8]{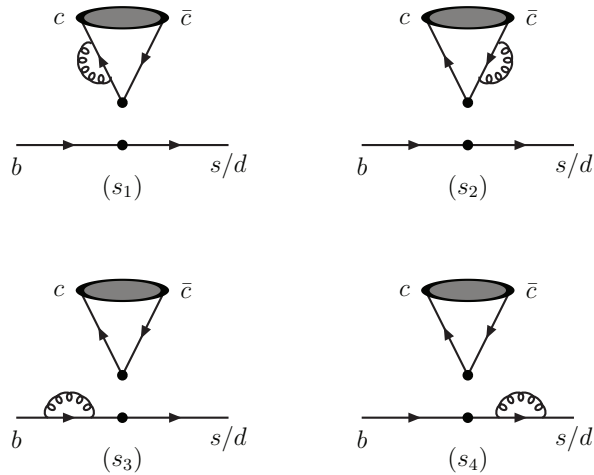}
\caption{Self-energy correction Feynman diagrams of $b\rightarrow
c\bar{c}+X$.} \label{figure:selfenergy}
\end{center}
\end{figure}

Adding self-energy diagrams in Fig.~[\ref{figure:selfenergy}], one can
remove UV divergences in $(v_3)$ and $(v_6)$.
Explicitly,
\begin{eqnarray}
(v_3) + (s_1) + (s_2) &=& \text{UV finite},\nonumber\\
(v_6) + (s_3) + (s_4) &=& \text{UV finite},
\end{eqnarray}
where
\begin{eqnarray}
(s_1)&=& -\frac{4}{3} \, i \, (4 \pi \alpha_s) N_{\epsilon}(m_b)  \left[-\frac{1}{2
   \epsilon_{UV}}+\frac{3}{2} \log \left(\frac{\eta }{4}\right)-\log (\xi )-2\right](tree),\nonumber\\
(s_2)&=& -\frac{4}{3} \, i \, (4 \pi \alpha_s) N_{\epsilon}(m_b)  \left[-\frac{1}{2
   \epsilon_{UV}}+\frac{3}{2} \log \left(\frac{\eta }{4}\right)-\log (\xi )-2\right](tree),\nonumber\\
(s_3)&=& -\frac{4}{3} \, i \, (4 \pi \alpha_s) N_{\epsilon}(m_b)  \left[-\frac{1}{2
  \epsilon_{UV}}-\log (\xi )-2\right] (tree) ,\nonumber\\
(s_4)&=& -\frac{4}{3} \, i \, (4 \pi \alpha_s) N_{\epsilon}(m_b)  \left[-\frac{1}{2
   \epsilon_{UV}}+\frac{\log (\xi )}{2}+\frac{1}{4}\right](tree),
\end{eqnarray}
with $N_{\epsilon}(m_b)=i \, (4 \pi )^{\epsilon_{UV}-2} \Gamma
(\epsilon_{UV}+1) \left(\frac{\mu
^2}{m_b^2}\right)^{\epsilon_{UV}}$. No virtual corrections to
${}^1P_1^{[8]}$ and ${}^1D_2^{[1]}$ exist accurate to NLO in
$\alpha_s$, because of their vanishing tree-level amplitudes. This
leads to a convenience directly that computation is reduced
significantly. $(v_1)+(v_2)+(v_4)+(v_5)$ is still UV divergent,
which needs operator renormalization, {\it i.e.}, to subtract the
term proportional to $\frac{1}{\epsilon_{UV}}-\gamma_E+\ln(4\pi)$ or
equivalently make the replacement
\begin{equation}
\frac{1}{\epsilon_{UV}}\rightarrow \gamma_E-\ln(4\pi).
\end{equation}
$\gamma_E$ is the Euler constant. To summarize our renormalization procedures.
First, make mass renormalization for charm, anti-charm and bottom quarks
$m_R\rightarrow m_0=m_R+m_{ct}$ (No such operation is needed for strange
or down quarks which are taken to be massless in this paper.),
\begin{equation}
m_{ct}=\frac{4}{3} \, i \, (4\pi\alpha_s)  N_{\epsilon}(m_b)\left[\frac{3}{\epsilon_{UV}}+4\right]m_R \, ;
\end{equation}
second, add the self-energy diagrams of external quark lines;
finally, do operator renormalization explained above.

\subsection{Residual Divergence Cancellation}

\begin{figure}[htbp]
\begin{center}
\includegraphics[scale=0.8]{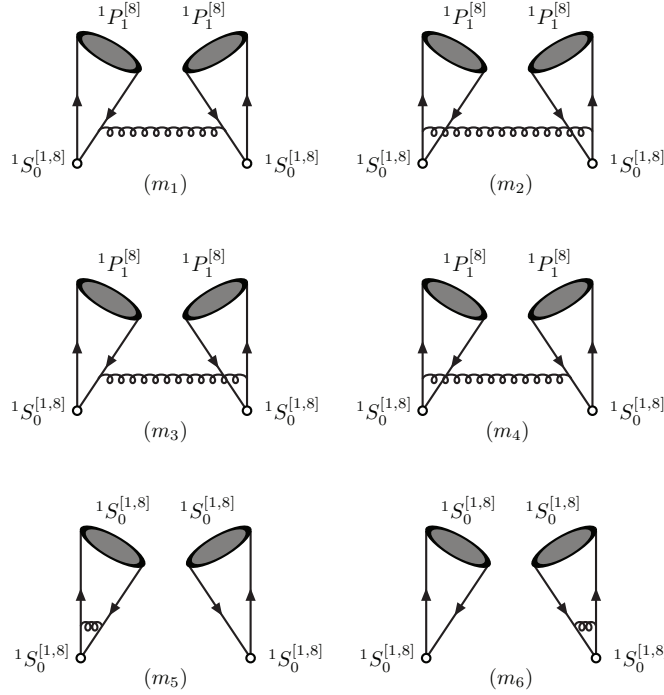}
\caption{NRQCD operator mixing of ${}^1S_0^{[1,8]}$ and ${}^1P_1^{[8]}$.}
\label{figure:matrixelements}
\end{center}
\end{figure}

We then demonstrate how the residual IR divergence is cancelled. At
NLO in $\alpha_s$, on the QCD side,
\begin{eqnarray}\label{matching:QCD}
\Gamma(b\rightarrow \eta_{c2}X)&=&C({}^1S_0^{[1]})^{QCD}_{finite+Coulomb}
\langle\mathcal{O}_1({}^1S_0)\rangle_{Born}+ C({}^1S_0^{[8]})^{QCD}_{finite+Coulomb}
\langle\mathcal{O}_8({}^1S_0)\rangle_{Born}\nonumber\\
&+& C({}^1P_1^{[8]})^{QCD}_{soft}\frac{\langle\mathcal{O}_8({}^1P_1)
\rangle_{Born}}{m_c^2}+ C({}^1D_2^{[1]})^{QCD}_{finite}
\frac{\langle\mathcal{O}_1({}^1D_2)\rangle_{Born}}{m_c^4},
\end{eqnarray}
while on the NRQCD side,
\begin{eqnarray}
\Gamma(b\rightarrow \eta_{c2}X)&=& C({}^1S_0^{[1]})^{NR}
\langle\mathcal{O}_1({}^1S_0)\rangle^{NR}+ C({}^1S_0^{[8]})^{NR}
\langle\mathcal{O}_8({}^1S_0)\rangle^{NR}\nonumber\\
&+& C({}^1P_1^{[8]})^{NR}\frac{\langle\mathcal{O}_8({}^1P_1)
\rangle^{NR}}{m_c^2}+ C({}^1D_2^{[1]})^{NR}\frac{\langle\mathcal{O}_1({}^1D_2)
\rangle^{NR}}{m_c^4}\nonumber.
\end{eqnarray}
The subscript $Coulomb$ or $soft $ means having Coulomb or soft
pole. NRQCD operator mixing of ${}^1S_0^{[1,8]}$ and ${}^1P_1^{[8]}$
is shown in Fig.~[\ref{figure:matrixelements}]. Similar for
${}^1P_1^{[8]}$ mixing with ${}^1D_2^{[1]}$.
And the non-perturbative matrix elements up to NLO in $\alpha_s$ are
\begin{eqnarray}\label{eq:mu-dependence}
\langle\mathcal{O}_1({}^1S_0)\rangle^{NR}&=&\langle\mathcal{O}_1({}^1S_0)\rangle_{Born}+
\langle\mathcal{O}_1({}^1S_0)\rangle_{Coulomb}-\frac{\alpha_s}{4\pi}
(\ln\frac{\lambda^2}{\mu_\Lambda^2}+\frac{1}{3})(\frac{16}{3})
\frac{\langle\mathcal{O}_8({}^1P_1)\rangle_{Born}}{m_c^2},\nonumber\\
\langle\mathcal{O}_8({}^1S_0)\rangle^{NR}&=&\langle\mathcal{O}_8({}^1S_0)
\rangle_{Born}+\langle\mathcal{O}_8({}^1S_0)\rangle_{Coulomb}\nonumber\\
&-&\frac{\alpha_s}{4\pi}(\ln\frac{\lambda^2}{\mu_\Lambda^2}+\frac{1}{3})(\frac{16}{3})
\left(C_F\frac{\langle\mathcal{O}_1({}^1P_1)\rangle_{Born}}{2N_cm_c^2}+B_F
\frac{\langle\mathcal{O}_8({}^1P_1)\rangle_{Born}}{m_c^2}\right),\nonumber\\
\langle\mathcal{O}_8({}^1P_1)\rangle^{NR}&=&\langle\mathcal{O}_8({}^1P_1)
\rangle_{Born}+\langle\mathcal{O}_8({}^1P_1)\rangle_{Coulomb}\nonumber\\
&-&\frac{\alpha_s}{4\pi}(\ln\frac{\lambda^2}{\mu_\Lambda^2}+\frac{1}{3})(\frac{16}{3})
\left(C_F\frac{\langle\mathcal{O}_1({}^1D_2)\rangle_{Born}}{2N_cm_c^2}
+B_F\frac{\langle\mathcal{O}_8({}^1D_2)\rangle_{Born}}{m_c^2}\right).
\end{eqnarray}
$B_F=\frac{5}{12}$. The Coulomb singularity in $(m_5)$ and $(m_6)$
of Fig.~[\ref{figure:matrixelements}] is extracted and related to
the tree-level matrix element in the following way
\begin{equation}
\langle\mathcal{O}_{[n]}(c\bar{c})\rangle_{Coulomb}=\mathcal{C}_{[n]}\frac{\pi\alpha_s}{2v}
\langle\mathcal{O}_{[n]}(c\bar{c})\rangle_{Born},
\end{equation}
with the color factor
\begin{equation}
\mathcal{C}_{[n]}=\begin{cases}
C_F=\frac{4}{3},& \text{n=1\,\,\, color-singlet $c\bar{c}$}\,;\\
-\frac{1}{2N_c}=-\frac{1}{6},& \text{n=8\,\,\, color-octet $c\bar{c}$}\,,
\end{cases}
\end{equation}
leading to
\begin{eqnarray}\label{matching:NRQCD}
\Gamma(b\rightarrow \eta_{c2}X)&=& C({}^1S_0^{[1]})^{NR}\langle\mathcal{O}_1({}^1S_0)
\rangle_{Born}+ C({}^1S_0^{[1]})_{Born}\langle\mathcal{O}_1({}^1S_0)\rangle_{Coulomb}\nonumber\\
&-& C({}^1S_0^{[1]})_{Born}\frac{\alpha_s}{4\pi}(\ln\frac{\lambda^2}
{\mu_\Lambda^2}+\frac{1}{3})(\frac{16}{3})\frac{\langle\mathcal{O}_8({}^1P_1)
\rangle_{Born}}{m_c^2}\nonumber\\
&+& C({}^1S_0^{[8]})^{NR}\langle\mathcal{O}_8({}^1S_0)\rangle_{Born}+ C({}^1S_0^{[8]})_{Born}
\langle\mathcal{O}_8({}^1S_0)\rangle_{Coulomb}\nonumber\\
&-& C({}^1S_0^{[8]})_{Born}\frac{\alpha_s}{4\pi}(\ln\frac{\lambda^2}
{\mu_\Lambda^2}+\frac{1}{3})(\frac{16}{3})B_F\frac{\langle\mathcal{O}_8({}^1P_1)
\rangle_{Born}}{m_c^2}\nonumber\\
&+& C({}^1P_1^{[8]})^{NR}\frac{\langle\mathcal{O}_8({}^1P_1)
\rangle_{Born}}{m_c^2}+ C({}^1D_2^{[1]})^{NR}\frac{\langle\mathcal{O}_1({}^1D_2)
\rangle_{Born}}{m_c^4}.
\end{eqnarray}
Matching Eq.~(\ref{matching:QCD}) and Eq.~(\ref{matching:NRQCD}),
one can get the finite short-distance coefficients accurate to
one-loop level
\begin{eqnarray}
C({}^1S_0^{[1]})^{NR}&=& C({}^1S_0^{[1]})^{QCD}_{finite}\,,\nonumber\\
C({}^1S_0^{[8]})^{NR}&=& C({}^1S_0^{[8]})^{QCD}_{finite}\,,\nonumber\\
C({}^1P_1^{[8]})^{NR}&=& C({}^1P_1^{[8]})^{QCD}_{soft}\nonumber\\
&+& C({}^1S_0^{[1]})_{Born}\frac{\alpha_s}{4\pi}(\ln\frac{\lambda^2}
{\mu_\Lambda^2}+\frac{1}{3})(\frac{16}{3})\nonumber\\
&+& C({}^1S_0^{[8]})_{Born}\frac{\alpha_s}{4\pi}(\ln\frac{\lambda^2}
{\mu_\Lambda^2}+\frac{1}{3})(\frac{16}{3})B_F\,,\nonumber\\
C({}^1D_2^{[1]})^{NR}&=& C({}^1D_2^{[1]})^{QCD}_{finite}\,.
\end{eqnarray}
Coulomb singularities in $C({}^1S_0^{[1]})^{QCD}$ and $C({}^1S_0^{[8]})^{QCD}$
and soft divergence in $C({}^1P_1^{[8]})^{QCD}$ are absorbed into the
long-distance matrix elements $\langle\mathcal{O}_1({}^1S_0)\rangle^{NR}$ and
$\langle\mathcal{O}_8({}^1S_0)\rangle^{NR}$. There is no residual soft divergence
in real correction to ${}^1D_2^{[1]}$ because of the absence of tree-level amplitude
of ${}^1P_1^{[8]}$. Considering its vanishing virtual correction, the NLO correction
to ${}^1D_2^{[1]}$ is finite. One-loop level short-distance coefficient can
be expressed in the common form
\begin{equation}\label{width:oneloop}
C(b\rightarrow c\bar{c}[n]+x)=\Gamma_0\frac{\alpha_s}{4\pi}
(C^2_{[1]}g_1[n]+2C_{[1]}C_{[8]}g_2[n]+C^2_{[8]}g_3[n]),
\end{equation}
and $g_1[n]$, $g_2[n]$ and $g_3[n]$ of ${}^1S_0^{[1]}$,
${}^1S_0^{[8]}$ and ${}^1P_1^{[8]}$ were calculated in
\cite{Beneke:1998ks,Maltoni:1999phd}. We list them in Appendix. B.
For ${}^1D_2^{[1]}$, our results are new:
\begin{eqnarray}
g_1[{}^1D_2^{[1]}]&=&0,\nonumber\\
g_2[{}^1D_2^{[1]}]&=&0,\nonumber\\
g_3[{}^1D_2^{[1]}]&=&\frac{8}{135} \left(2 \eta ^3-9 \eta ^2+18 \eta -6 \log (\eta )-11\right).
\end{eqnarray}

\subsection{Evaluation of long-distance matrix elements}
Due to lack of experimental information on the matrix elements of
D-wave operators, we can not extract them from experiments and have
to invoke some theoretical estimates. The color-singlet matrix
element $\langle\mathcal{O}_1({}^1D_2)\rangle$ may be determined by
potential models with input parameters, while the color-octet matrix
elements may be estimated using the operator evolution equations.
Matrix elements $\langle\mathcal{O}_8({}^1P_1)\rangle^{NR}$,
$\langle\mathcal{O}_1({}^1S_0)\rangle^{NR}$ and
$\langle\mathcal{O}_8({}^1S_0)\rangle^{NR}$ are renormalized in
NRQCD, and thus have $\mu_{\Lambda}$-dependence, and this can be
explicitly shown by deriving the quantities on both sides of
Eq.~(\ref{eq:mu-dependence}) with respect to $\mu_{\Lambda}$:
\begin{eqnarray}\label{eq:evolution} \frac{d
\langle\mathcal{O}_1({}^1S_0)\rangle^{NR}}{d\ln\mu_\Lambda}&=&\frac{\alpha_s}
{4\pi}\frac{32}{3}\frac{\langle\mathcal{O}_8({}^1P_1)\rangle_{Born}}{m_c^2},
\nonumber\\
\frac{d
\langle\mathcal{O}_8({}^1S_0)\rangle^{NR}}{d\ln\mu_\Lambda}&=&
\frac{\alpha_s}{4\pi}\frac{32}{3}B_F\frac{\langle\mathcal{O}_8({}^1P_1)\rangle_{Born}}{m_c^2},
\nonumber\\
\frac{d
\langle\mathcal{O}_8({}^1P_1)\rangle^{NR}}{d\ln\mu_\Lambda}&=&
\frac{\alpha_s}{4\pi}\frac{32}{3}C_F\frac{\langle\mathcal{O}_1({}^1D_2)\rangle_{Born}}{2N_cm_c^2}.
\end{eqnarray} Eq.~(\ref{eq:evolution}) has the same form as Eq.~(45)
in \cite{Fan:2009cj}, where the IR divergence is regularized in
dimensional regularization scheme. This is because the operator
evolution equations have nothing to do with the IR divergent parts.
The solutions are
\begin{eqnarray}\label{solution:matrixelementsEvolution}
\langle\mathcal{O}_8({}^1P_1)(\mu_\Lambda)\rangle^{NR}&=&
\frac{1}{2N_c}\frac{8C_F}{3m_c^2b_0}\ln\frac{\alpha_s(\mu_{\Lambda_0})}
{\alpha_s(\mu_{\Lambda})}\langle\mathcal{O}_1({}^1D_2)\rangle_{Born},
\nonumber\\
\langle\mathcal{O}_1({}^1S_0)(\mu_\Lambda)\rangle^{NR}&=&\frac{1}{2N_c}\frac{C_F}{2}
\Big(\frac{8}{3m_c^2b_0}\ln\frac{\alpha_s(\mu_{\Lambda_0})}{\alpha_s(\mu_{\Lambda})}
\Big)^2\langle\mathcal{O}_1({}^1D_2)\rangle_{Born},
\nonumber\\
\langle\mathcal{O}_8({}^1S_0)(\mu_\Lambda)\rangle^{NR}&=&\frac{1}{2N_c}\frac{C_FB_F}{2}
\Big(\frac{8}{3m_c^2b_0}\ln\frac{\alpha_s(\mu_{\Lambda_0})}
{\alpha_s(\mu_\Lambda)}\Big)^2\langle\mathcal{O}_1({}^1D_2)\rangle_{Born},
\end{eqnarray}
where we take $m_c=1.5$\,GeV, $b_0=\frac{11C_A}{6}-\frac{N_f}{3}$,
$C_A=3$, $N_f$=3, $\Lambda^{LO}_{QCD}=153$\,MeV for LO, and
$\Lambda_{QCD}=399$\,MeV for NLO.

The initial matrix elements like
$\langle\mathcal{O}_8({}^1P_1)(\mu_{\Lambda_0})\rangle$ at starting
scale $\mu_{\Lambda_0}=m_c v$, where $v^2=0.25$, are eliminated. One
could refer to \cite{Fan:2009cj} for reasonability of doing so. The
evolution equation method for determining the long-distance matrix
elements has been used in estimating the D-wave charmonium state
light hadronic decay width and $h_c$ decay
width\cite{Fan:2009cj,He:2009bf,He:2008xb,Fan:2010huyu}. For $h_c$,
the evolution equation could give a prediction for light hadronic
decay width within about 30\% error when compared to experimental
extraction\cite{Fan:2010huyu}. That means the operator evolution
equation is a good method to evaluate the P-wave long-distance
matrix element, and can be extended to D-wave case, which is lack of
experimental data.


\section{results and discussions}
\label{sec:results} The long-distance CS D-wave matrix element is
related to the second derivative of the radial wave function at the
origin
\begin{equation}
\langle\mathcal{O}_1(n{}^1D_2)\rangle=(2J+1)\langle n {}^1D_2|\mathcal{O}_1(n{}^1D_2)|n {}^1D_2\rangle=5\,
(2N_c)\,\frac{15|R^{\prime\prime}_{nD}(0)|^2}{8\pi},
\end{equation}
where $N_c=3$ and B-T potential model input parameter $|R^{\prime\prime}_{1D}(0)|^2=0.015$ GeV$^7$
\cite{Eichten:1995ch} for charmonium. Before giving the final results, we have to first deal
with the NLO Wilson coefficients $C_{[1]}(\mu)$ and $C_{[8]}(\mu)$.
The expressions for $C_\pm(\mu)$ up to NLO in $\alpha_s$ are given in \cite{Buras:1989xd}
\begin{equation}
C_\pm(\mu) = \left[ \frac{\alpha_s(M_W)}{\alpha_s(\mu)} \right]
^{\gamma^{(0)}_\pm/(2 \beta_0)}
\left( 1 + \frac{\alpha_s(\mu)}{4\pi} B_\pm  \right)
\left( 1 + \frac{\alpha_s(M_W)-\alpha_s(\mu)}{4\pi} (B_\pm-J_\pm) \right),
\end{equation}
with
\begin{eqnarray}
J_\pm &=&\frac{\gamma^{(0)}_\pm\beta_1}{2 \beta_0^2} -
\frac{\gamma^{(1)}_\pm}{2 \beta_0},
\nonumber\\
B_\pm &=& \frac{3\mp 1}{6} (\pm 11 + \kappa_\pm),
\end{eqnarray}
and the one-loop and two-loop anomalous dimensions
\begin{eqnarray}
\gamma^{(0)}_\pm &=& \pm \,2 \,(3 \mp 1),
\nonumber\\
\gamma^{(1)}_\pm &=& \frac{3 \mp 1}{6} \left(-21\pm\frac{4}{3}\,N_f
- 2 \beta_0 \kappa_\pm \right).
\end{eqnarray}
The scheme-dependent $\kappa_\pm$ are
\begin{equation}
\kappa_\pm=\begin{cases}
0,& \text{NDR scheme},\\
\mp 4,& \text{HV scheme}.
\end{cases}
\end{equation}
Note here an additional factor $-\frac{16}{3}$ should be included in
$B_\pm$ in the HV scheme.  $\beta_0$ and $\beta_1$ are in the NLO
expression for $\alpha_s$
\begin{equation}
\alpha_s(\mu) = \frac{4 \pi}{\beta_0 \ln (\mu^2/\Lambda_{QCD}^2)}
\left[ 1 - \frac{\beta_1 \ln [\ln (\mu^2/\Lambda_{QCD}^2)]}
{\beta_0^2 \ln (\mu^2/\Lambda_{QCD}^2)} \right],
\end{equation}
with $\Lambda_{QCD}=$ 345 MeV, $\beta_0 = 11 - \frac{2}{3} \, N_f$,
and $\beta_1 = 102 - \frac{38}{3} \, N_f $.

\begin{figure}[htbp]
\begin{center}
\includegraphics[scale=0.6]{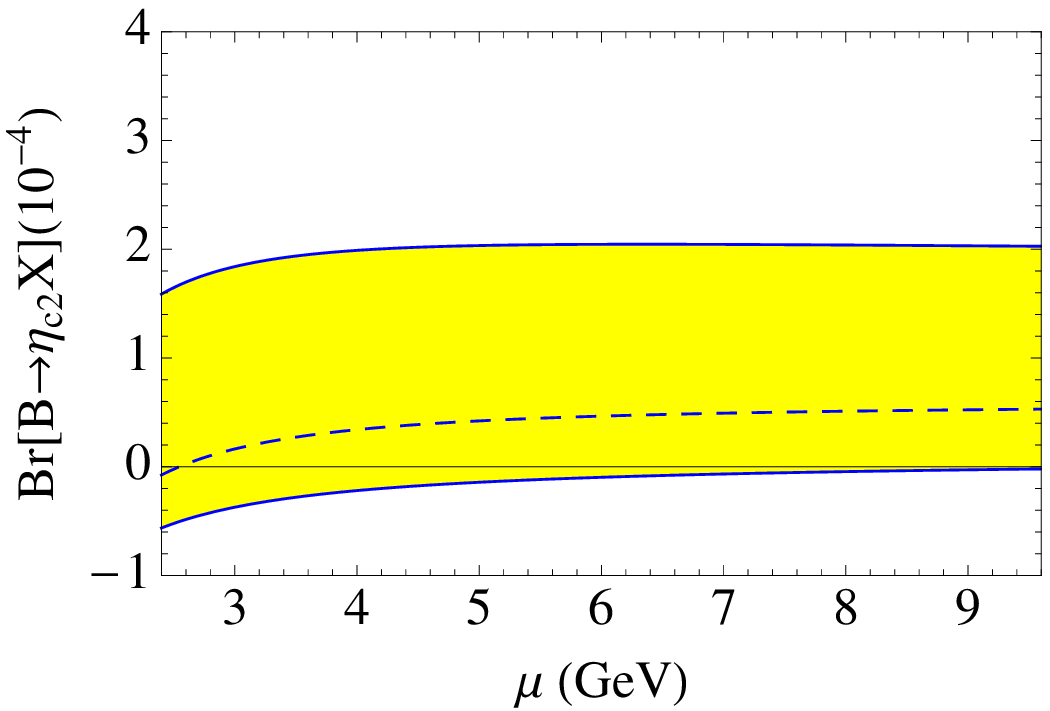}
\hspace{1cm}
\includegraphics[scale=0.6]{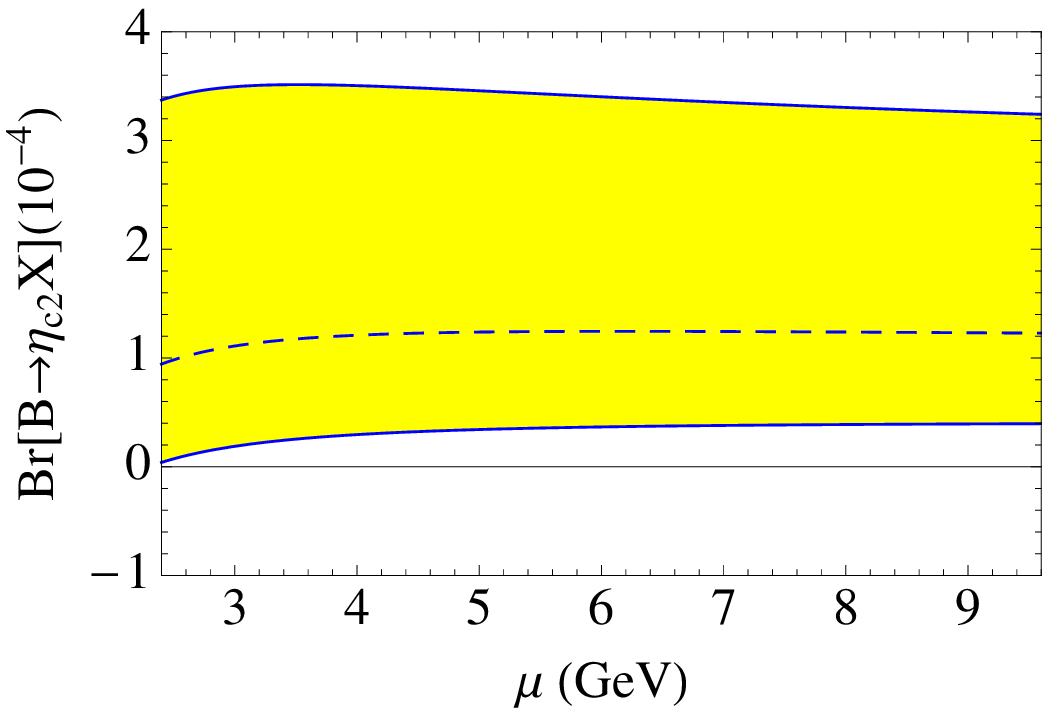}
\caption{QCD renormalization scale $\mu$-dependence of
$Br[B\rightarrow \eta_{c2}X]$ in  NDR scheme (left) and HV scheme
(right). The long-distance matrix elements are estimated using
operator evolution equations.
$\mu$ ranges from $\frac{m_b}{2}$ to $2m_b$. The shaded zone is for
the values of $Br[B\rightarrow \eta_{c2}X]$. Upper bound for solid
curves correspond to $m_c=1.4$\,GeV and $\mu_{\Lambda_0}=700$\,MeV,
dashed lines to $m_c=1.5$\,GeV and $\mu_{\Lambda_0}=750$\,MeV, and
lower bound solid curves to $m_c=1.6$\,GeV and
$\mu_{\Lambda_0}=800$\,MeV, respectively.} \label{figure:muBr}
\end{center}
\end{figure}

LO and NLO short-distance contributions are given in Table
\ref{table:shortdistance}. It is easy to see that at renormalization
scale $\mu=m_b$, the short-distance coefficients in  NDR and HV
schemes differ slightly for the dominant components ${}^1P_1^{[8]}$
and ${}^1S_0^{[8]}$.
\begin{table}[t]
\addtolength{\arraycolsep}{0.15cm}
\renewcommand{\arraystretch}{1.2}
\caption[]{\label{table:shortdistance} LO
(Eq.\,(\ref{width:LOwithPenguin})) and NLO (both
Eq.\,(\ref{width:LOwithPenguin}) and Eq.\,(\ref{width:oneloop}))
short-distance coeffecients of four subprocesses, with $\Gamma_0$
removed. Results for both NDR and HV schemes are listed. The QCD
renormalization scale $\mu$ takes values from $\frac{m_b}{2}$ to
$2m_b$, where $m_b=4.8$\,GeV, $m_c=1.5$\,GeV. }
$$
\begin{array}{c|c|c|c|c|c|c|c|c|c}
\hline
\hline
\mbox{Fock state} & \multicolumn{3}{|c|}{\mbox{LO}} & \multicolumn{3}{|c}{\mbox{NLO NDR scheme}} &
\multicolumn{3}{|c}{\mbox{NLO HV scheme}}\\
\hline
\mu & m_b/2 & m_b & 2m_b & m_b/2 & m_b & 2m_b & m_b/2 & m_b & 2m_b \\
\hline
{}^1D_2^{[1]}  &   0   &  0  &  0   & 0.0028 & 0.0020 & 0.0015 & 0.0026 & 0.0018 & 0.0014 \\
{}^1P_1^{[8]}  &   0   &  0  &  0   & -2.058 & -1.545 & -1.289 & -1.880 & -1.390 & -1.150 \\
{}^1S_0^{[1]}  & 0.0458 & 0.2130 & 0.4330 & -0.2102 & -0.3978 & -0.4892 & -0.0633 & -0.1950 & -0.2629  \\
{}^1S_0^{[8]}  &  8.803 & 8.065 & 7.566 & 12.856 & 11.217 & 10.169 & 13.490 & 11.529 & 10.287  \\
\hline
\hline
\end{array}
\vspace*{0.2cm}
$$
\end{table}
The long-distance matrix elements take the following values
\begin{eqnarray}\label{eq:matrixvalueevolution}
\frac{\langle\mathcal{O}_1({}^1D_2)\rangle}{m_c^4}&=&0.053~\mbox{GeV$^3$},\nonumber\\
\frac{\langle\mathcal{O}_8({}^1P_1)\rangle}{m_c^2}&=&0.0092~\mbox{GeV$^3$},\nonumber\\
\langle\mathcal{O}_1({}^1S_0)\rangle&=&0.0036~\mbox{GeV$^3$},\nonumber\\
\langle\mathcal{O}_8({}^1S_0)\rangle&=&0.0015~\mbox{GeV$^3$},
\end{eqnarray}
where $m_c=1.5$\,GeV and $\mu_{\Lambda_0}=m_c v=750$\,MeV. The
long-distance matrix elements
$\frac{\langle\mathcal{O}_8({}^1P_1)\rangle}{m_c^2}$,
$\langle\mathcal{O}_1({}^1S_0)\rangle$ and
$\langle\mathcal{O}_8({}^1S_0)\rangle$ are sensitive to charm quark
mass $m_c$ and initial scale $\mu_{\Lambda_0}$. Multiplying the
short-distance coefficients shown in Table \ref{table:shortdistance}
by the matrix elements in Eq.~(\ref{eq:matrixvalueevolution}), we
get the B-meson semi-inclusive decay width into $\eta_{c2}$. Then we
can estimate its branching ratio using B-meson inclusive
semi-leptonic decay rate. That has the benefit of eliminating the
$V_{bc}$ dependence and reducing the $m_b$ dependence, as was
performed in
\cite{Bergstrom:1994vc,Soares:1997ir,Maltoni:1999phd,Beneke:1998ks}.
The theoretical prediction for the inclusive semi-leptonic decay
width can be expressed as\cite{Altarelli:1991dx}
\begin{equation}\label{width:SL}
\Gamma_{SL}=\frac{G_F^2|V_{bc}|^2 m_b^5}{192 \pi^3}\,[1-8z^2+8z^6-z^8-24z^4\log(z)]\,\eta_1(z),
\end{equation}
where $z=\frac{m_c}{m_b}$. The factor $\eta_1(z)$, including NLO QCD
correction, has the approximate form \cite{Kim:1989ac}
\begin{equation}
\eta_1(z)=1-\frac{2\alpha_s(\mu)}{3\pi}\left[\frac{3}{2}+(-\frac{31}{4}+\pi^2)(1-z)^2\right].
\end{equation}
Using the calculated B-meson semi-inclusive decay width given in
Eq.~(\ref{width:SL}), and the experimental semi-leptonic branching
ratio $Br_{SL}=10.74\%$ \cite{Nakamura:2010zzi}, and taking $m_c$
and $\mu_{\Lambda_0}$ in regions (1.4,\,1.6)GeV and (700,\,800)MeV,
respectively, we finally arrive at the QCD renormalization scale
$\mu$-dependence curves in Fig.~[\ref{figure:muBr}] for the
branching ratio $Br[B\rightarrow \eta_{c2}X]$ of B-meson
semi-inclusive decay into $\eta_{c2}$. Note that varying
$\mu_{\Lambda_0}$ only changes the relative ratios among
long-distance matrix elements, while varying $m_c$ affects not only
the long-distance matrix elements but also the short-distance
coefficients.

When $\mu$ is taken to be $m_b=4.8$\,GeV,
\begin{eqnarray}\label{eq:Br}
Br(B\rightarrow \eta_{c2}X)_{NDR}&=&(\,0.41^{+1.62}_{-0.56}\,)\,\times10^{-4},\nonumber\\
Br(B\rightarrow \eta_{c2}X)_{HV}&=&(\,1.24^{+2.23}_{-0.90}\,)\,\times10^{-4},
\end{eqnarray}
where the central values correspond to $m_c=1.5$\,GeV and
$\mu_{\Lambda_0}=750$\,MeV, upper bounds to $m_c=1.4$\,GeV and
$\mu_{\Lambda_0}=700$\,MeV, and lower bounds to $m_c=1.6$\,GeV and
$\mu_{\Lambda_0}=800$\,MeV, respectively. Since the color-octet
Wilson coefficient $C_{[8]}(\mu)$ is much larger than the
color-singlet one $C_{[1]}(\mu)$
\begin{equation}
\frac{C^2_{[8]}(\mu)}{C^2_{[1]}(\mu)}\approx 15\,,
\end{equation}
the LO decay width is dominated by that of ${}^1S_0^{[8]}$, which is
proportional to $C^2_{[8]}(\mu)$. For NLO, decay widths of
${}^1S_0^{[1]}$ and ${}^1D_2^{[1]}$ are negligible, and those of
${}^1P_1^{[8]}$ and ${}^1S_0^{[8]}$ are of the same order and make
most contribution to the branching ratio in Eq.~(\ref{eq:Br}), but
unluckily they largely cancel each other. This cancellation is
related to our estimates for the long-distance matrix elements in
Eq.~(\ref{eq:matrixvalueevolution}). If without this cancellation,
the ${}^1S_0^{[8]}$ Fock state could give the following central
values
\begin{eqnarray}
Br(B\rightarrow {}^1S_0^{[8]}X)_{NDR}&=&\,5.30\,\times10^{-4},\nonumber\\
Br(B\rightarrow {}^1S_0^{[8]}X)_{HV}&=&\,5.45\,\times10^{-4},
\end{eqnarray}
which might be regarded as the upper bound of the branching ratio
for this process.  Furthermore, we may consider the following
uncertainty in the predictions of the branching ratio. Since
\begin{equation}
\frac{C_{[1]}}{C_{[8]}}\sim \alpha_s\,,
\end{equation}
we might carry out a double expansion in both $\alpha_s$ and
$C_{[1]}/C_{[8]}$ simultaneously \cite{Bergstrom:1994vc}. In this
new expansion, terms of different orders scale as follows:
\begin{align}
\text{LO:} & \quad C^2_{[8]}\,; \nonumber\\
\text{NLO:} & \quad \alpha_s C^2_{[8]},\; C_{[1]} C_{[8]}  \,; \nonumber\\
\text{N$^2$LO:} & \quad \alpha_s^2 C^2_{[8]}, \; \alpha_s C_{[1]} C_{[8]}, \; C^2_{[1]} \,; \nonumber\\
\text{N$^3$LO:} & \quad \alpha_s^3 C^2_{[8]}, \; \alpha_s^2 C_{[1]}
C_{[8]}, \; \alpha_s C^2_{[1]}, \cdots \,.
\end{align}
$C^2_{[8]}$ scales as LO, and $\alpha_s C^2_{[8]}$ as NLO.
$\alpha_s^2 C^2_{[8]}$ scales the same order as $\alpha_s C_{[1]}
C_{[8]}$ and $C^2_{[1]}$, thus should also be considered. Authors of
\cite{Bergstrom:1994vc} did not calculate all $\alpha_s^2 C^2_{[8]}$
terms, but estimated their contribution by adding a correction term
of the same order. The same method with a minor modification was
adopted in \cite{Maltoni:1999phd,Beneke:1998ks}. Unluckily, their
method can only be applied to the color-singlet channels that have
non-vanishing LO decay widths, and fails in our case. In
\cite{Soares:1997ir}  the $\alpha_s^2 C^2_{[8]}$ virtual
contribution from squared one-loop amplitudes was calculated, but
the real correction was neglected by arguing that the real
contribution was phase-space suppressed. However, the IR divergent
real corrections can not be omitted, as pointed out in
\cite{Maltoni:1999phd,Beneke:1998ks}. Hence, a complete calculation
at NNLO in $\alpha_s$ might be needed to obtain the $\alpha_s^2
C^2_{[8]}$ contribution, but this is already beyond the scope of our
calculation in this paper. It will be interesting to see if the
large cancellation of ${}^1P_1^{[8]}$ and ${}^1S_0^{[8]}$ could be
weakened after including the $\alpha_s^2 C^2_{[8]}$ contribution.

We now discuss the possible relation between the semi-inclusive
decay branching ratio $B\rightarrow \eta_{c2} X$ and the exclusive
decay branching ratio $B\rightarrow \eta_{c2} K$. Obviously, the
latter must be much smaller than the former, since the $X$ includes
many hadronic states other than the kaon. In particular, in the case
of $B\rightarrow \eta_{c2} X$, the dominant contribution comes from
the color-octet $c\bar c$ channels, which subsequently evolve into
$\eta_{c2}$ by emitting soft gluons which then turn into light
hadrons such as pions. Whereas the exclusive process $B\rightarrow
\eta_{c2} K$ requires the soft gluons be reabsorbed by the strange
quark in $b\to c\bar c+s$. This probability is apparently very
small. As a conservative estimate, we believe the branching ratio of
$B\rightarrow \eta_{c2} K$ should be smaller than that of
$B\rightarrow \eta_{c2} X$ by at least an order of magnitude. The
suppression of exclusive decay relative to inclusive decay is
supported by many other charmonium states. E.g., the branching ratio
of $B\rightarrow J/\psi X$ is $(7.8\pm0.4)\times 10^{-3}$
\cite{Nakamura:2010zzi}, while $Br(B^+\rightarrow J/\psi
K^+)=(1.014\pm0.034)\times 10^{-3}$ and $Br(B^0\rightarrow J/\psi
K^0)=(8.71\pm0.32)\times 10^{-4}$. For $\chi_{c1}$, $Br(B\rightarrow
\chi_{c1} X)=(3.22\pm0.25)\times 10^{-3}$, $Br(B^+\rightarrow
\chi_{c1} K^+)=(4.6\pm0.4)\times 10^{-4}$ and $Br(B^0\rightarrow
\chi_{c1} K^0)=(3.90\pm0.33)\times 10^{-4}$. Evidently, the observed
inclusive branching ratios are about 10 times larger than the
corresponding exclusive one. For $\chi_{c2}$, which is similar to
$\eta_{c2}$ because in both cases at LO the color-singlet $c\bar c$
Fock states make no contributions, $Br(B\rightarrow
\chi_{c2}X)=(1.65\pm0.31)\times 10^{-3}$, $Br( B^+\rightarrow
\chi_{c2}K^+)<1.8\times 10^{-5}$ and $Br(B^0\rightarrow
\chi_{c2}K^0)<2.6\times 10^{-5}$, the suppression of exclusive decay
is almost by two-order of magnitude. Therefore, we may have a
general observation that for a charmonium state produced in B-meson
decays, the suppression factor of exclusive production branching
ratio relative to inclusive one should not be larger than 1/10
(including the factorizable and non-factorizable exclusive
processes). This means $Br(B\rightarrow \eta_{c2}K)$ should be at
most $\mathcal{O}(10^{-5})$, based on our calculation.

In contrast, for X(3872) the observed branching ratio
$Br(B\rightarrow X(3872) K)\times Br(X(3872)\rightarrow
D^0\bar{D}^0\pi^0)=(1.2\pm0.4)\times 10^{-4}$
\cite{Nakamura:2010zzi}. Considering that there exist many decay
modes of X(3872) other than $X(3872)\to D^0\bar{D}^0\pi^0$, we may
conclude that $Br(B\rightarrow X(3872) K)$ is at least 10 times
larger than $Br(B\rightarrow \eta_{c2}K)$. Therefore, $X(3872)$ is
unlikely to be the $J^{PC}=2^{-+}$ charmonium state $\eta_{c2}$. In
fact, for X(3872) the $J^{PC}=1^{++}$ assignments of the
$D^0\bar{D}^{*0}$ molecule\cite{Tornqvist:X3872:molecule} or a
charmonium-$D^0\bar{D}^{*0}$ mixed
state\cite{Meng:2005er,Suzuki:2005ha} are preferred by many authors,
instead of a $J^{PC}=2^{-+}$ state (for more discussions see a
recent review \cite{Brambilla:2010cs}).

\section{conclusions}
In this paper, we calculate the semi-inclusive decay width and
branching ratio of $B\rightarrow \eta_{c2}X$ at NLO in $\alpha_s$ in
NRQCD factorization framework. The finite short-distance
coefficients are obtained by matching QCD and NRQCD, and the
non-perturbative long-distance matrix elements are evaluated by
using the operator evolution equations. We find that at tree-level,
only the S-wave Fock states ${}^1S_0^{[1,8]}$ contribute, and the LO
decay width is dominated by that of ${}^1S_0^{[8]}$, because of the
largeness of the color-octet Wilson coefficient squared
$C^2_{[8]}(\mu)$ over the color-singlet one $C^2_{[1]}(\mu)$. Unlike
$\eta_{c2}$ light hadronic decay, in this process, there is no
residual divergence at NLO of the ${}^1D_2^{[1]}$ Fock state, due to
the vanishing tree-level contribution of ${}^1P_1^{[8]}$. At NLO in
$\alpha_s$, ${}^1P_1^{[8]}$ and ${}^1S_0^{[8]}$ dominate.
Unfortunately, they largely cancel each other. This cancellation
depends on our method for estimating the long-distance matrix
elements. As a result, we obtain the branching ratio
$Br(B\rightarrow
\eta_{c2}X)=(\,0.41^{+1.62}_{-0.56}\,)\,\times10^{-4}$ in the NDR
scheme and $(\,1.24^{+2.23}_{-0.90}\,)\,\times10^{-4}$ in the HV
scheme, at $\mu=m_b$. The central values correspond to
$m_c=1.5$\,GeV and $\mu_{\Lambda_0}=750$\,MeV, upper bounds to
$m_c=1.4$\,GeV and $\mu_{\Lambda_0}=700$\,MeV, and lower bounds to
$m_c=1.6$\,GeV and $\mu_{\Lambda_0}=800$\,MeV, respectively. If the
large cancellation does not exist, the ${}^1S_0^{[8]}$ could give
$Br(B\rightarrow {}^1S_0^{[8]}X)_{NDR}=\,5.30\,\times10^{-4}$ and
$Br(B\rightarrow {}^1S_0^{[8]}X)_{HV}=\,5.45\,\times10^{-4}$, which
could be regarded as the upper bound of the branching ratio of this
process. The $\mu$-dependence curves of NLO branching ratios in the
two schemes are also shown, where $\mu$ varies from $\frac{m_b}{2}$
to $2m_b$ and $\mu_{\Lambda}=2m_c$. Furthermore, we estimate the
exclusive decay branching ratio of $B\rightarrow \eta_{c2} K$ by
considering the suppression ratios of exclusive decays relative to
inclusive ones for other factorizable and non-factorizable exclusive
charmonium production processes, and conclude that $X(3872)$ is
unlikely to be a $2^{-+}$ charmonium state. We hope that our results
will be useful in finding the missing charmonium state $\eta_{c2}$
in experiments, and in further studying $\eta_{c2}$ production in
B-meson exclusive decays.

\section{acknowledgments}

We would like to thank Yan-Qing Ma and Yu-Jie Zhang for many helpful
discussions. This work was supported by the National Natural Science
Foundation of China (Nos.10721063, 11021092, 11075002), the Ministry
of Science and Technology of China (No.2009CB825200), and the  China
Postdoctoral Science Foundation (No.2010047010).

\section{Appendix}
\subsection{Covariant projector method.}
In our calculation of short-distance coefficients, the covariant
projector method is adopted\cite{Keung:1982jb}. For any spin-singlet
charmonium production in 4-dimension, the covariant projector is
\begin{equation}
\bar{P}_{0,0}(P,k)=\frac{1}{2\sqrt{2}}\frac{\slashed{p}_3-m_c}{\sqrt{\frac{M}{2}+m_c}}\gamma^5
\frac{\slashed{P}+M}{M}\frac{\slashed{p}_4+m_c}{\sqrt{\frac{M}{2}+m_c}}\,,
\end{equation}
where momentum of charmonium bound state $P=p_4+p_3$. Relative
momentum between charm quark and anti-charm quark satisfies
\begin{eqnarray}
p_4&=&\frac{P}{2}+k,
\nonumber\\
p_3&=&\frac{P}{2}-k.
\end{eqnarray}
Bound state mass $M=2m_c$, which holds in QCD radiative correction
calculations, for the relativistic effects are neglected. For more
details, one could refer to related contents in \cite{Fan:2009cj}.

\subsection{One-loop level short-distance coefficients of
${}^1S_0^{[1]}$, ${}^1S_0^{[8]}$ and ${}^1P_1^{[8]}$ Fock states.}
For ${}^1S_0^{[1]}$,
\begin{eqnarray}
g_1[{}^1S_0^{[1]}]&=&-4 (1-\eta )^2 \left(8 \text{Li}_2(\eta )-4 \text{Z}+4 \log (1-\eta ) \log (\eta )+\frac{4 \pi
   ^2}{3}\right)\nonumber\\
&+&20 (1-\eta )^2+\frac{8 (2-5 \eta ) (1-\eta )^2 \log (1-\eta )}{\eta }-16 \eta
   (1-\eta ) \log (\eta ),\nonumber\\
g_2[{}^1S_0^{[1]}]&=&4 (1-\eta )^2 \left(3 \log \left(\frac{m_b^2}{\mu
   ^2}\right)-\text{X}+\text{Y}\right)-\frac{2 \left(17 \eta ^2-53 \eta +34\right) (1-\eta
   )}{2-\eta }\nonumber\\
&+&4 \eta ^2 \log (\eta )+\frac{8 (3-\eta ) (1-\eta )^3 \log (1-\eta )}{(2-\eta )^2},\nonumber\\
g_3[{}^1S_0^{[1]}]&=&\frac{4}{9} \left(-(1-\eta ) \left(2 \eta ^2-7 \eta +11\right)-6 \log (\eta )\right);
\end{eqnarray}

for ${}^1S_0^{[8]}$,
\begin{eqnarray}
g_1[{}^1S_0^{[8]}]&=&-\frac{4}{3} (1-\eta ) \left(2 \eta ^2-7 \eta +11\right)-8 \log (\eta ),\nonumber\\
g_2[{}^1S_0^{[8]}]&=&3 (1-\eta )^2 \left(3 \log \left(\frac{m_b^2}{\mu
   ^2}\right)-\text{X}+\text{Y}\right)-\frac{3 \left(17 \eta ^2-53 \eta +34\right) (1-\eta
   )}{2 (2-\eta )}
   \nonumber\\
&&+3 \eta ^2 \log (\eta )+\frac{6 (3-\eta ) (1-\eta )^3 \log (1-\eta )}{(2-\eta
   )^2},\nonumber\\
g_3[{}^1S_0^{[8]}]&=&\frac{9}{2} (1-\eta )^2 \Bigg(-4 \log \left(\frac{m_b^2}{\mu ^2}\right)+\frac{4
   \text{X}}{3}+\frac{14 \text{Y}}{3}-\frac{2 \text{Z}}{3}
   -3 \log ^2(2-\eta )\nonumber\\
   &&
   +6 \log (1-\eta) \log (2-\eta )-6 \log (2) \Bigg)
   +3 (1-\eta ) \Bigg(9 (\eta +1)
   \text{Li}_2\left(\frac{1-\eta }{2-\eta }\right)
   \nonumber\\
   &&
   -18 \text{Li}_2\left(\frac{2 (1-\eta )}{2-\eta
   }\right)+(7 \eta +29) \text{Li}_2(\eta )-\frac{1}{6} \pi ^2 (29 \eta +7)
   \nonumber\\
   &&
   +18 \log (2) \log
   (2-\eta )+2 (4 \eta +5) \log (1-\eta ) \log (\eta )-18 \log (2) \log (\eta
   ) \Bigg)
   \nonumber\\
   &&+\frac{1}{2} \left(90 \eta ^2-48 \eta +17\right) \log (\eta )+\frac{\left(20 \eta
   ^3+2077 \eta ^2-6221 \eta +4478\right) (1-\eta )}{12 (2-\eta )}
   \nonumber\\
   &&
   -\frac{3 \left(33 \eta ^3-113
   \eta ^2+106 \eta +4\right) (1-\eta )^2 \log (1-\eta )}{(2-\eta )^2 \eta };
\end{eqnarray}

and for ${}^1P_1^{[8]}$,
\begin{eqnarray}
g_1[{}^1P_1^{[8]}]&=&16 (1-\eta )^2 \left(2 \log (1-\eta )-\log \left(\frac{\mu_\Lambda ^2}{4
  m_c^2}\right)\right)-\frac{4}{9} \left(8 \eta ^2-85 \eta +119\right) (1-\eta
   )\nonumber\\
   &&-\frac{8}{3} \left(12 \eta ^2-6 \eta +1\right) \log (\eta ),\nonumber\\
g_2[{}^1P_1^{[8]}]&=&0,\nonumber\\
g_3[{}^1P_1^{[8]}]&=&10 (1-\eta )^2 \left(2 \log (1-\eta )-\log \left(\frac{\mu_\Lambda ^2}{4
  m_c^2}\right)\right)-\frac{1}{9} \left(29 \eta ^2-244 \eta +347\right) (1-\eta
   )\nonumber\\
   &&-\frac{2}{3} \left(30 \eta ^2-15 \eta +7\right) \log (\eta ).
\end{eqnarray}

\end{document}